\documentclass[aps,twocolumn,floats,superscriptaddress,showpacs]{revtex4}

\usepackage{color}
\usepackage{graphicx}

\begin{document}

\title{Mott-Hubbard Scenario for the Metal-Insulator Transition in the
Two Dimensional Electron Gas}

\author{Ping Sun}
\affiliation{Department of Physics and Astronomy, Rutgers University,
Piscataway, NJ 08854-8019}

\date{\today}

\begin{abstract}
By comparing the responses to an in-plane magnetic field near the
metal-insulator transition (MIT), we find that the observed MIT in Si
MOSFETs can be described by the non-perturbative Mott-Hubbard
scenario. Interrelations between independent measurables are uncovered
and confirmed by replotting the experimental data. A universal
critical energy scale vanishing at the MIT is extracted from the
experimental data and the critical exponent found.
\end{abstract}


\maketitle

The properties of an electron gas (EG) are controlled by the average
inter-electron distance. As the electron density is tuned from high to
low, the Coulomb interaction, favoring localization of the electrons,
becomes dominant over the kinetic energy and the system evolves from a
metal to an insulator\cite{wigner}. Puzzling magnetotransport
experiments on the silicon metal-oxide-semiconductor field-effect
transistors (Si MOSFETs) in the past ten years\cite{abra,sara} have
reactivated broad interest in the subject because of possible
relations to the detection of a quantum critical point (QCP), the
realization of the Wigner crystal\cite{wigner}, and the interplay
between correlation and disorder\cite{gang4,fink,cast,punn}. By
comparing the responses to an in-plane magnetic field near the
metal-insulator transition (MIT), we find that the observed MIT is
controlled by the non-perturbative Mott-Hubbard (M-H) critical end
point at a low but non-zero temperature\cite{mott,hubbard,rmp96},
instead of a QCP. This also contradicts the long-standing speculation
that the non-local part of the Coulomb interaction plays an essential
role in the MIT, as it does in the Wigner crystalization\cite{wigner}.
The M-H scenario predicts interrelations between independent
measurables which are confirmed by replotting the experimental
data. The uncovered sample independent behavior indicates that the
disorder does not play an active role. A critical energy scale is
extracted from the experimental data and the critical exponent found.

An MIT in the two dimensional (2D) EG was first observed via the
conductivity measurements in ultra clean Si MOSFETs\cite{kra94}. When
the temperature is decreased below $\sim 2 K$, the conductivity is
observed to increase (decrease) monotonically at densities above
(below) a certain non-universal critical value $n_c\sim 10^{11}
cm^{-2}$.  In this temperature range, the conductivity curves in a
given phase can be all collapsed into a single curve by rescaling the
temperature\cite{kra95}. The observation clearly points to an
unidentified type of critical behavior. It was also found\cite{kra95}
that the scaling behavior is sample independent down to the lowest
temperatures reached experimentally ($\lesssim 100mK$). The most
likely driving force is then the interaction induced correlation,
instead of the disorder.

In the subsequent experiments in Si MOSFETs, it was found that near
$n_c$ applying an increasing in-plane magnetic field $H_{||}$ causes
the conductivity first to decrease and then, above a certain field
$H_{\sigma}$, saturate at a roughly field independent
value\cite{vit01,kra01}. As the temperature is lowered, $H_{\sigma}$
first decreases linearly and then approaches a constant. This constant
depends on the electron density and vanishes as the transition is
approached from the metallic side\cite{vit01}. At $H_{||} >
H_{\sigma}$, the conductivity is always insulating like, decreasing as
the temperature is lowered, irrespective of whether the density
corresponds to a metallic or an insulating phase in the absence of the
field\cite{shas01}. Shubnikov-de Haas (SdH) measurements\cite{vit00}
show that the oscillation period is halved when $H_{||} \gtrsim
H_{\sigma}$. This has been taken as an indication that the entire EG
is fully polarized when the conductivity saturates in the
field\cite{kra01}. Since $H_{\sigma}$ vanishes at $n=n_c$ and $T=0$,
this could point to a Stoner instability\cite{ston}, although it would
be hard to understand the almost coincidence of the Stoner instability
with the MIT\cite{kra01}. Direct\cite{rez,kra06} and indirect
\cite{vit01,kra01,puda02} measurements show that the homogeneous spin
susceptibility is enhanced in the approach to the transition and is
related to a mass enhancement since the g-factor changes
little\cite{shas02}.

The perturbative scaling theory for non-interacting disordered
electron systems shows that no true metallic phase exist in 2D due to
the disorder\cite{gang4}. However, the Coulomb interaction competes
with the localization and stablizes the metallic
phase\cite{fink,cast,punn}. The relevance of the scaling theory to the
experiments was checked\cite{cast}. Starting from the Fermi liquid,
perturbations were used to study the temperature dependence of the
conductivity\cite{hwang,zala} and the strong mass
enhancement\cite{sarma}. While they describe the 2DEG to certain
extent, these perturbations are not justified since the
experiments\cite{abra,sara} on Si MOSFETs show that the transition
happens when $R_S$, the average inter electron distance measured in
units of the Bohr radius, is $\gtrsim 10$. The significance of this is
that $R_S$ is also given by the ratio of the interaction energy, {\it
i.e.} the perturbation, to the kinetic energy.

The only non-perturbative, correlation-driven MIT we understand is the
M-H transition\cite{mott,hubbard,rmp96}. The M-H scenario was related
to the 2DEG before\cite{spiv,carol}. However, the clear connections
between the two become evident only through studying the
magnetotransport behavior. Within this scenario, we manage to explain
the experiments and uncover previously unnoticed interrelations
between independent observables.

The M-H transition, driven by the ratio of the on-site interaction $U$
to the bandwidth $W$, is best revealed by the behavior of the local
density of states (LDOS) as shown by the dynamical mean field theory
(DMFT)\cite{rmp96}. Near the transition and on the metallic side, the
LDOS has a three peak structure (Fig. 1), with a central peak at the
Fermi level and two side bands separated in energy by roughly $U$. The
former consists of extended quasiparticle states; its width is
proportional to the quasiparticle weight $Z$. The latter, the upper
and lower Hubbard bands, consist of localized states. In between the
peaks, the LDOS is featureless, corresponding to an incoherent
background. As one moves towards the insulating side by increasing
$U/W$, the width of the central peak decreases to zero. DMFT
found\cite{rmp96} that, irrespective of the bare dispersion details,
the MIT happens at $U/(W/2) \sim 3$ and is first-order at zero
temperature. The first-order transition line extends to non-zero
temperatures and ends in a critical point at $T_c\sim 0.05\cdot (W/2)$
beyond which the transition becomes a crossover. This picture was
confirmed experimentally only recently, {\it e.g.} in the
photoemission studies\cite{mo} of V$_2$O$_3$ at $T \gtrsim T_c$. At
$T\ll T_c$ the physics is usually non-universal and depends on the
system details.

We argue that the magnetotransport experiments on the 2DEG can be
explained by the evolution in a magnetic field of the three peak LDOS.
On the metallic side, an increasing in-plane magnetic field first
splits the quasiparticle peak while reducing its height. This reduces
the LDOS near the Fermi surface and thus the conductivity. When the
field is strong enough so that the quasiparticle peak is either
entirely split apart or suppressed, the LDOS left over near the Fermi
surface derives only from the incoherent background. Consequently the
conductivity saturates at a value determined by the short mean free
time of those states and its temperature dependence becomes
insulating-like. The field at which this saturation appears is
$H_{\sigma} \propto (1/g\mu_B) \cdot {\rm max\;}(ZW, k_BT)$, meaning
the Zeeman energy should be strong enough to split or suppress the
quasiparticle peak and overcome the thermal excitations. So as $T$ is
reduced, $H_{\sigma}$ first decreases linearly and then saturates to a
constant $\propto Z$. Since $Z$ reduces to zero as the transition is
reached from the metallic side, $H_{\sigma}(T=0)$ vanishes
accordingly. Meanwhile, the effective mass ($\propto 1/Z$) gets
enhanced and diverges.

To support the above explanation, we have computed the evolution in an
in-plane magnetic field of the LDOS near the M-H transition using a
half-filled lattice Hubbard model,

\[
   H=\sum_{\sigma=\pm}
   \int_{|\vec{k}|\le \Lambda} \frac{d^2k}{(2\pi)^2}
   \left(\epsilon_k + \sigma H_{\parallel} \right)
   C^{\dag}_{\vec{k},\sigma} C_{\vec{k},\sigma}
\]
\[
   + U \sum_{i} N_{i\uparrow} N_{i\downarrow}
\]

\noindent where we used $N_{i,\sigma}=C^{\dag}_{i,\sigma}C_{i,\sigma}$
and set the $g$-factor $= 2$. We approximate the first Brillouin zone
by a disk. The momentum cut-off $\Lambda=2\sqrt{\pi}$ ensures the
normalization of the momentum integral and sets the lattice constant
to be $2\pi/\Lambda$. We use the free dispersion $\epsilon_k=k^2/2$
specific to the EG. The bandwidth is thus $W=2\pi$. This effective
model is valid when the thermal and Zeeman energies are small
comparing to the bandwidth. We mention, though, even at low
temperatures there is a non-zero possibility that an electron jumps to
a higher energy state. In terms of the one-band lattice model, this is
like an electron hopping away from the lattice sites and into the
interstitials.

In this dimensionless model for the 2DEG, the energy is measured in
units of $E_0=2\epsilon_BR_s^{-2}$, where
$\epsilon_B=m_be^4/(2\epsilon^2 \hbar^2)$, with the electron band mass
$m_b\sim 0.2 m_e$ and the dielectric constant for the Si MOSFET
$\epsilon \sim 7.7$. Since in the experiments, the MIT is observed
around $R_s \sim 10$, we estimate $E_0$ to be $\sim 10.6 K$. The
corresponding magnetic field unit is, using $g=2$, $H_0=E_0/(\mu_B
m_e/m_b) \sim 3.1 T$. A rough estimation of the non-universal $T_c$
for the M-H critical end point is possible at this stage. Applying the
DMFT estimation in the current model, we obtain $T_c \sim 1.7 K$. The
mean field estimation is about one order of magnitude higher than the
experimental value $T_c \lesssim 100 mK$.

To compute the conductivity, we use the Kubo formula\cite{rmp96}. In
the EG, the current operator is given by $j(\vec{q}) \sim
\sum_{\vec{k},\sigma} (\vec{k}+\vec{q}/2)
C^{\dag}_{\vec{k}+\vec{q},\sigma}C_{\vec{k},\sigma}$. Only the
particle-hoel bubble survives in the current-current correlation in
DMFT \cite{rmp96}. The conductivity, in unit of $e^2/(\hbar d)$,

\[
   \sigma = 
   \lim_{\omega\rightarrow 0} \frac{1}{\omega}
   \sum_{\sigma=\pm}
   \int_{|\vec{k}|\le \Lambda} \frac{d^2k}{(2\pi)^2}
   \frac{1}{\beta} \sum_{ip_n} 
\]
\[
   k^2 G_{\sigma}(\vec{k},ip_n)
   G_{\sigma}(\vec{k},ip_n+ i \omega),
\]

\noindent with $p_n=(2n+1)\pi/\beta$.

\begin{figure}[ht]
\label{fig-dos}
\includegraphics[width=7cm]{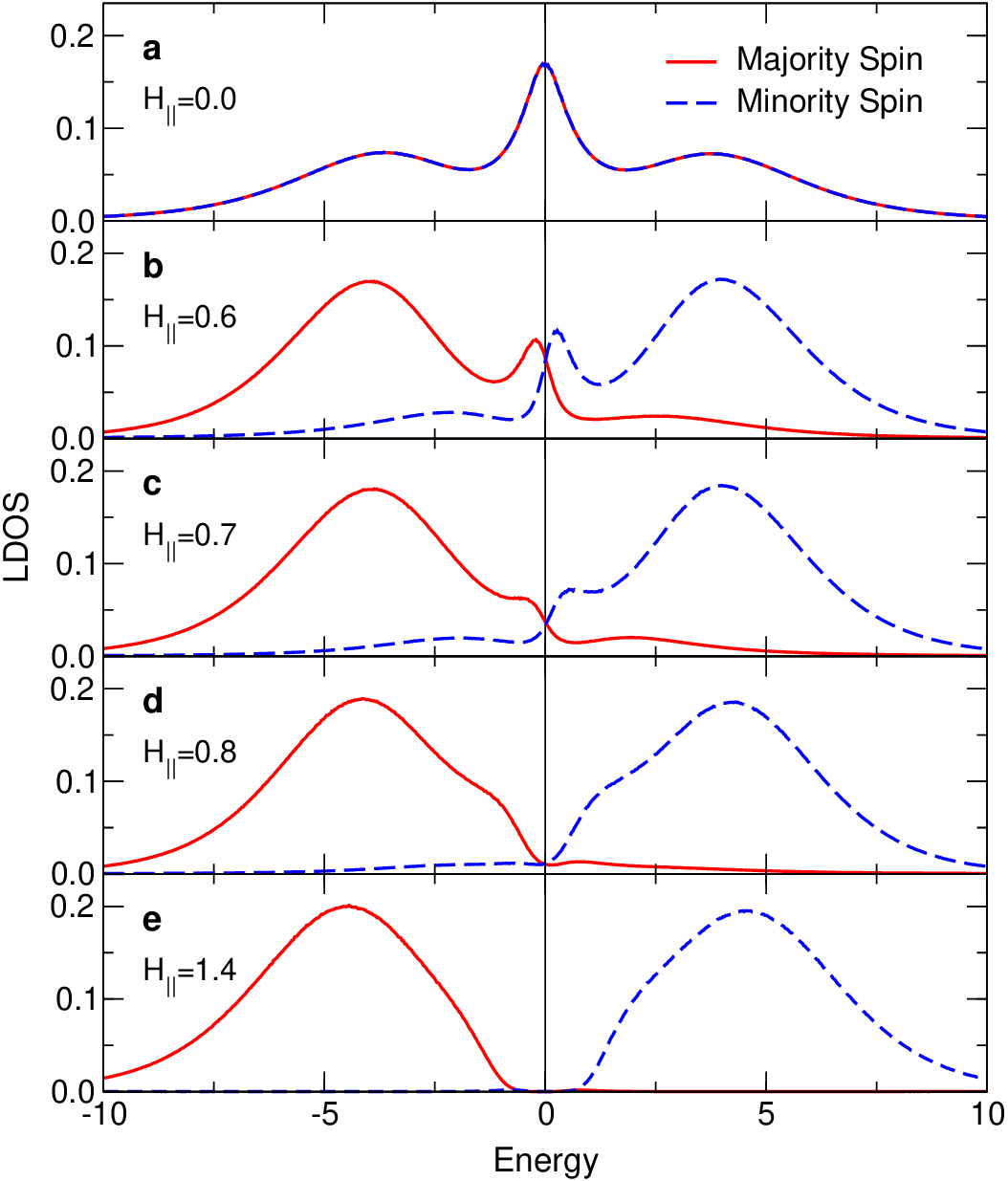}
\caption{The evolution of the LDOS in an external in-plane magnetic
field $H_{||}$. At zero field, the LDOS shows a quasiparticle peak at
the Fermi energy ($E=0$) together with two side bands. An increasing
$H_{||}$ causes the peak to split and finally get suppressed. At
$H_{||} \sim 0.7$ when the quasiparticle peak is fully split, the LDOS
shows that the localized spins are still not entirely polarized. In
this model calculation, the strong incoherent background as observed
experimentally in the M-H systems\cite{mo} is absent from the LDOS.}
\end{figure}

\begin{figure}[ht]
\label{fig-cnd}
\includegraphics[width=7cm]{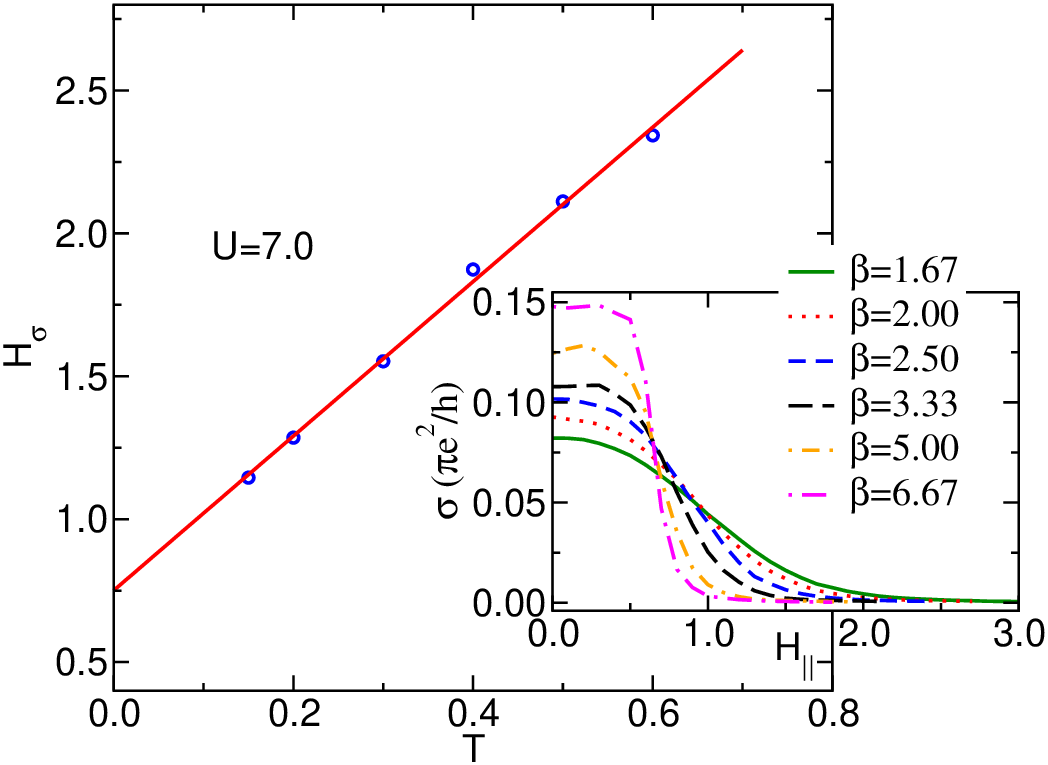}
\caption{The conductivity calculated via DMFT\cite{rmp96} as a
function of the magnetic field at different temperatures is shown in
the inset. When the field is strong enough, the conductivity
saturates. The saturation field $H_{\sigma}$ is determined when the
conductivity reaches $\sigma=2 \times 10^{-3} \; \pi e^2/h$, which is
about the order of the QMC statistical error. (The choice of this
value does not change the qualitative behavior of $H_{\sigma}$.)
$H_{\sigma}$ {\it vs} $T$ is shown in the main plot by the circles,
together with a linear fitting by the line. The intersection of the
line at $T=0$ gives an estimation $ZW\sim 0.75$ at $U=7.0$.}
\end{figure}

We solve this model numerically via DMFT, using quantum Monte Carlo
(QMC) as the impurity solver\cite{rmp96}.  In Fig. 1 the evolution of
the LDOS is presented at an inverse temperature $\beta=5.0$ and an
on-site interaction $U=7.0$. The latter is chosen so that the system
is on the metallic side and close to the MIT. The conductivity
saturation behavior is shown in the inset of Fig. 2. The $T$ {\it vs}
$H_{\sigma}$ relation is shown in the main plot. In the temperature
region solved, $H_{\sigma}$ depends linearly on $T$, similar to that
observed experimentally\cite{vit01}.

Although it provides us with some essential features regarding the MIT
in 2DEG, this effective model contains certain limitations. These
include the lack of the incoherent hopping of the electrons into the
interstitial space. This should be responsible for the observed
saturated conductivity being non-zero at $T > 0$, instead of vanishing
as shown in the inset of Fig. 2. Actually it was found experimentally
that the saturated conductivity increases with the temperature and at
$T\gtrsim 1 K$ it can reach the same order of magnitude as that
measured at a metallic density and with no external
field\cite{shas01}. SdH measurements then become possible even at
$H_{||} \gtrsim H_{\sigma}$. The observed halving of the SdH
oscillation period when $H_{||}>H_{\sigma}$ is thus due to the full
polarization of the itinerant electrons, including those within the
quasiparticle peak and those involved in the interstitial hopping. The
electrons localized in the lower Hubbard band do not contribute to the
transport.

\begin{figure}[ht]
\label{fig-mas}
\includegraphics[width=7cm]{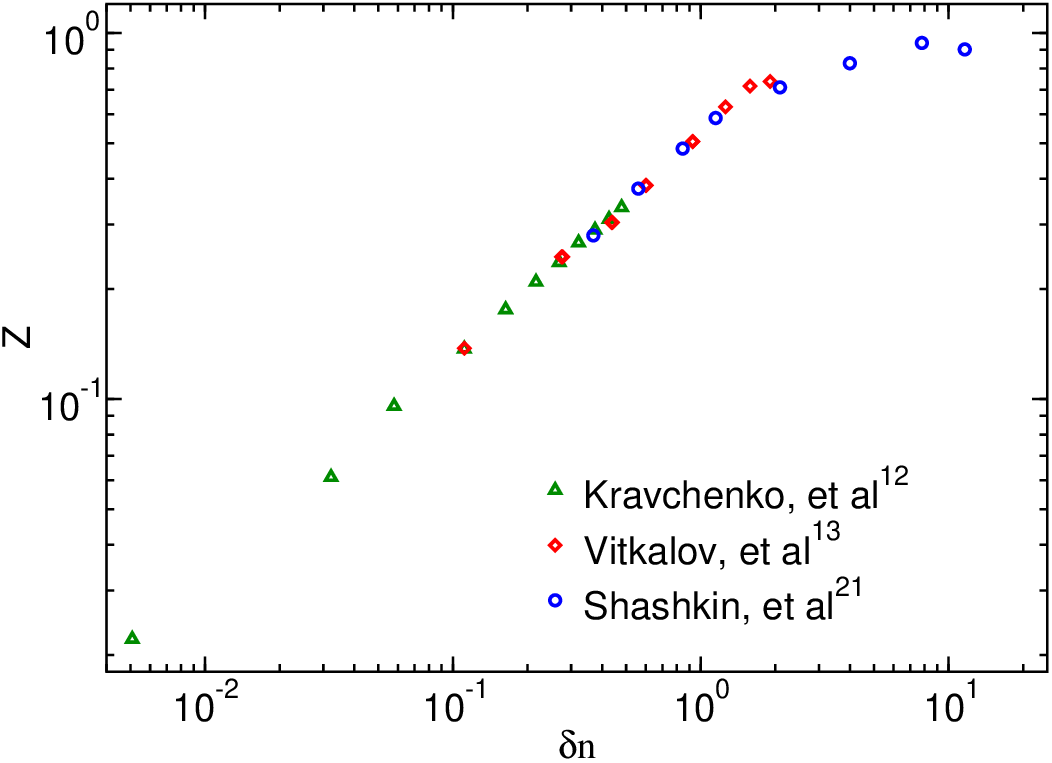}
\caption{Replotting of experimental data in terms of $Z$ {\it vs}
$\delta n = n/n_c -1$. Triangles correspond to the rescaling
temperature $T_0$ on the metallic side of sample Si-12a reported in
Ref.\cite{kra95}. It is transformed as $Z=T_0^{1/2.5}/10.5$. The
diamonds correspond to $H_{\sigma}(T=0)$ measured in
Ref.\cite{vit01}. It is rescaled as $Z=H_{\sigma}(T=0)/3.9$. The
circles represent the effective mass reported in Ref.\cite{shas02}
which is plotted as $Z=m_b/m^*$. The values of $n_c$ are taken from
the corresponding papers. Close to $\delta n = 0$, $Z$ can be fitted
to $Z\propto \delta n^{0.61}$. This relation is valid up to densities
almost three times the critical density.}
\end{figure}

While only one energy scale ($Z$) controls the critical transport
behavior, two distinct energy scales are responsible for the spin
polarization, the super exchange ($\propto 1/U$) for the local moments
and the quasiparticle peak width ($\propto Z$) the itinerant
electrons. Near the transition, $Z\rightarrow 0$ and it costs almost
no energy to polarize the itinerant electrons. Meanwhile, a non-zero
exchange energy must be overcome to polarize the local moments. This
behavior can be seen in the LDOS at $H\sim 0.7$ in Fig. 1, where the
quasiparticle peak is already fully split while the local moments are
only partially polarized. A direct proof of the two-energy-scale
behavior could be found from a measurement of the magnetization and a
study of its saturation behavior in a magnetic field\cite{lalo}. For
the 2DEG, no consensus has been reached
experimentally\cite{rez,kra06}, although indications favoring the
existence of the local moment band were reported\cite{rez}. This
convincing experiment would allow to discern the M-H scenario from the
Stoner picture. In the latter there is only one magnetic energy scale
and all the electrons are polarized simultaneously by a vanishing
field near the transition.

According to the M-H scenario, all the critical behaviors are related
to $Z$. The saturation magnetic field, $H_{\sigma}(T=0) \propto Z$
should vanish and the effective mass $m^* \propto 1/Z$ diverge in
approaching the transition from the metallic side. Besides, it is also
responsible for the vanishing energy scale $T_0$ as revealed by the
universal scaling of the conductivity\cite{kra95}, so $T_0\propto
Z^a$. In Fig. 3, we transformed, rescaled and replotted the
experimental scaling temperature $T_0$ data\cite{kra95}, the
$H_{\sigma}$ data\cite{vit01}, and the $m^*$
data\cite{shas02}. Although belonging to independent measurements on
different samples, the $H_{\sigma}$ and $m^*$ data sets collapse well
when their densities overlap, revealing the previously unnoticed
interrelations that we derived from the M-H scenario. From Fig. 3, we
find that $Z \propto \delta n^{0.61}$ for $\delta n \lesssim 2$. This
universal relation can be used to calibrate the critical
behaviors. Note that, the DMFT solution gives $Z\propto \delta n$,
since\cite{rmp96} $Z\propto U_c-U$ and the mapping from the density to
$U$ is expected to be analytic even around the transition. The
deviation of the critical exponent is not unexpected for a mean field
theory.

The M-H picture captures the universal critical behavior when $T_c
\lesssim T \lesssim 2K$, where the conductivity scaling was
achieved\cite{kra95} and all the other above-mentioned behaviors were
observed\cite{abra,sara}. The non-universal properties specific to
the 2DEG at $T \ll T_c$ were previously studied. Monte Carlo
simulations\cite{ceper} showed, in between the paramagnetic metallic
and the Wigner crystal phases, there lies at least one other phase,
the polarized liquid phase. A similar result was obtained by studying
the free energy. It was shown that a direct transition is forbidden
and various intermediate phases are present\cite{kiv}. At $T \gtrsim
2K$ the system crosses over to another different behavior. It was
observed that the conductivity as a function of temperature is
insulating like\cite{kra95}, irrespective of the density being above
or below $n_c$. This is likely due to the Anderson localization behavior
\cite{gang4} before the correlation sets in and the M-H critical point
takes over at lower temperatures.

We have managed to establish that the non-perturbative M-H critical
point describe the observed MIT in 2DEG. Our theory provides a
framework for understanding the interrelations between the independent
measurables, which is confirmed by replotting the experimental data. A
universal energy scale vanishing at the transition is uncovered. In
the light of the M-H scenario, we suggest further experiments on the
2DEG of measuring the magnetization saturation behavior in an in-plane
magnetic field. This will allow to reveal the predicted
two-energy-scale behavior near the transition. Subjects remain to be
clarified including the proper descriptions of the incoherent hopping
into the interstitials and the ineffectiveness of the
non-local part of the partially screened Coulomb interaction. To
fulfill these purposes, a controlled derivation of an effective low
energy lattice model for the EG will be helpful.

{\it Acknowledgements} The author thanks E. Abrahams for many fruitful
and illuminating discussions and proof reading of the manuscript.  He
thanks M. E. Gershenson and S. A. Vitkalov for helpful discussions of
the experiments in Si MOSFETs. He gratefully acknowledges discussions
with M. H. Cohen, G. Kotliar, A. J. Millis, and M. P. Sarachik. This
research was supported by the National Science Foundation and by the
Center for Materials Theory at Rutgers University.

\end{document}